\documentclass[preprint,prl,aps,amsmath,amssymb,color,superscriptaddress]{revtex4}
\usepackage{stmaryrd}
\usepackage{amsmath}
\usepackage{amssymb}
\usepackage{graphicx}
\usepackage{dcolumn}
\usepackage{bm}
\usepackage{tabularx}
\begin{document}
\begin{titlepage}
\title{Half Excitonic Insulator: A Single-Spin Bose-Einstein Condensate}
\author{Zeyu Jiang}
\affiliation{State Key Laboratory of Low-Dimensional Quantum Physics and Collaborative Innovation Center of Quantum Matter, Department of Physics, Tsinghua University, Beijing 100084, China}
\author{Yuanchang Li}
\email{yuancli@bit.edu.cn}
\affiliation{Key Lab of advanced optoelectronic quantum architecture and measurement (MOE), and Advanced Research Institute of Multidisciplinary Science, Beijing Institute of Technology, Beijing 100081, China}
\author{Wenhui Duan}
\affiliation{State Key Laboratory of Low-Dimensional Quantum Physics and Collaborative Innovation Center of Quantum Matter, Department of Physics, Tsinghua University, Beijing 100084, China}
\affiliation{Institute for Advanced Study, Tsinghua University, Beijing 100084, China}
\author{Shengbai Zhang}
\affiliation{Department of Physics, Applied Physics and Astronomy, Rensselaer Polytechnic Institute, Troy, NY, 12180, USA}
\date{\today}

\begin{abstract}
First-principles calculations reveal an unusual electronic state (dubbed as half excitonic insulator) in monolayer 1$T$-$MX_2$ ($M$ = Co, Ni and $X$ = Cl, Br). Its one spin channel has a many-body ground state due to excitonic instability, while the other is characterized by a conventional band insulator gap. This disparity arises from a competition between the band gap and exciton binding energy, which exhibits a spin-dependence due to different orbital occupations. Such a state can be identified by optical absorption measurements and angle-resolved photoemission spectroscopy. Our theory not only provides new insights for the study of exciton condensation in magnetic materials but also suggests that strongly-correlated materials could be fertile candidates for excitonic insulators.
\end{abstract}

\maketitle
\draft
\vspace{2mm}
\end{titlepage}
Signal processing uses electrons whereas signal communication uses photons as the media\cite{High}. They together laid the foundation for today's information technology. The manipulation of electrons is based on a charge gap ($E_g$) in a semiconductor, while photons can be directly coupled to excitons, which are bound electron-hole pairs. Usually, exciton energies are inside $E_g$. However, there may be cases where exciton binding energy ($E_b$) exceeds $E_g$, in which a reconstructed excitonic insulator (EI) with a distinctive broken symmetry may form, as demonstrated by Kohn and coworkers\cite{Kohn} some half century ago. It subsequently acquires a many-body ground state characterized by the spontaneously-formed exciton condensate.\cite{Kogar,Rontani}

Since the proposal, people have kept searching for clues of EIs in semimetals and small-gap semiconductors. In comparison, less attention has been paid to strongly-correlated materials\cite{Kunes2015}, despite that the very existence of EI is the result of an electron-correlation effect, perhaps because of their relatively large $E_g$. In addition, the strongly-correlated materials with localized $d$- or $f$-electrons often exhibit magnetism(s). Spontaneous exciton condensation in such systems naturally involves interactions with the spin degree of freedom, in contrast to the spinless systems where only the charge degree of freedom matters. Despite the complexity, a study of the excitonic instability in magnetic materials may provide new physical insights exploiting the multi facets of charge gap, magnetism, and spontaneously-formed exciton condensate.

For simplicity, let us consider a magnetic system where spin-orbit coupling is small and hence, to a good approximation, can be ignored. For the sake of discussion, let us also define a transition energy $E_t = E_g - E_b$, at which a normal to excitonic insulator transition is energetically favored. In terms of $E_t$, there can be three qualitatively different physical systems as schematically illustrated in Fig. 1. (a) When both spin channels have $E_t > 0$, it is a magnetic band insulator; (b) when both spin channels have $E_t < 0$, it is an EI; and (c) when only one spin channel has $E_t < 0$, it is a new state of matter, dubbed here as a half excitonic insulator (HEI). The first two cases have been extensively studied in the literature but not the latter, which will be the focus here. First of all, both spin channels have gaps but with totally different physical origins, i.e., single-electron physics versus many-body physics. Second, as the excitons are electron-hole pairs, they naturally obey a bosonic statistics on a length scale longer than the exciton radius.\cite{Rontani} In contrast, electrons and holes in the other channel are charged fermions. Therefore, the two spin channels must respond differently to external stimulus, displaying simultaneously the characteristics of an EI and a band insulator. Third, as a consequence, coherent flow of excitons will take place in one spin channel, leading to a perfect insulator in terms of the charge and heat transports\cite{RontaniPRL,usEI}, while in the other channel a conventional dissipative transport can be expected. This would result in various single-spin-like behaviors like spin superfluid\cite{HanW} as the exciton channel contributes nothing to low-energy-excitation electronic processes, while the band insulator channel would not respond to stimuli that probe in particular the many-body phenomena. This unusual dual functionality could offer a broad and unprecedented application potential in electronics, spintronics, photonics, and beyond.

\begin{figure}[tbp]
\includegraphics[width=0.75\columnwidth]{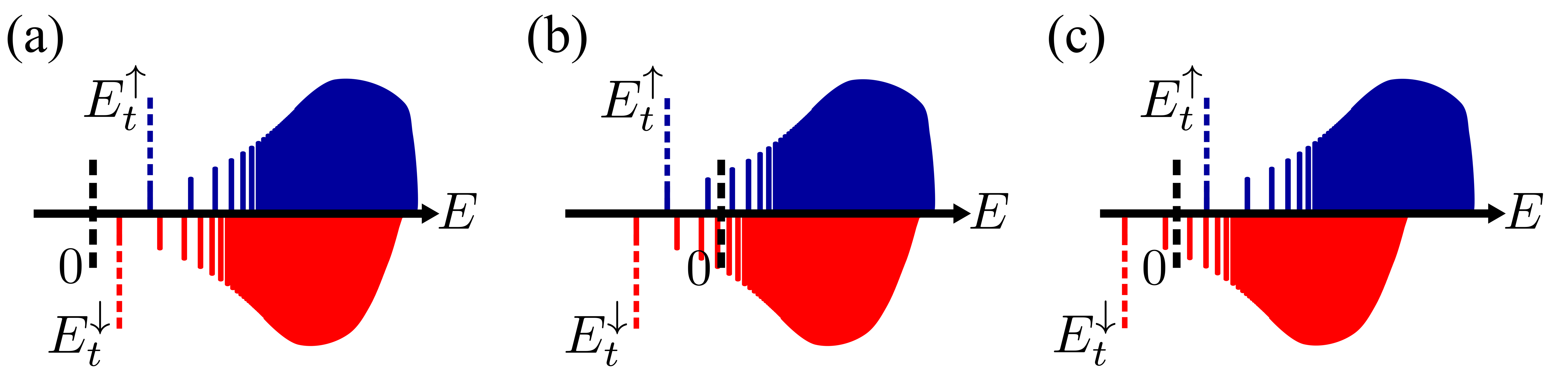}
\caption{\label{fig:fig1} (Color online) Schematic diagrams that classify magnetic insulators in terms of the exciton $E_t$: (a) an ordinary magnetic insulator when $E_t >0$ (both spins), (b) a magnetic EI when $E_t <0$ (both spins), and (c) a magnetic HEI when $E_t < 0$ (single spin). ``$\uparrow$" and ``$\downarrow$" denote electron spin. Thick black dashed lines denote the energy zero and the thin colored dashed lines denote the energetically most favorable excitons in each spin. Note that these magnetic systems can still be ferromagnetic or antiferromagnetic depending on their total magnetic moment.}
\end{figure}

In this paper, we show by first-principles calculations that HEI can indeed form in two-dimensional materials. We propose four of them for experimental verification. These are monolayer 1\emph{T}-$MX_2$ with $M$ = Co, Ni and $X$ = Cl, Br (whose parent forms are known to exist as layered materials \cite{McGuire}). We find $E_t$s for minority spin are all negative, i.e., $E_t =$ -83, -47, -590, and -540 meV for NiCl$_2$, NiBr$_2$, CoCl$_2$, and CoBr$_2$, but $E_t$s for majority spin are all positive. Interestingly, here $E_t$ can be larger than 20 $k_BT$ (at room temperature), which has been unimaginable before. This exceptionally large disparity between the spin channels can be traced back to the different characteristic low-energy $d$-$d$ and $d$-$sp$ orbital excitations that manifest themselves into a spin-dependent $E_b$ and $E_g$. We will discuss possible consequences due to the unique electronic structure of HEIs, as well as the key features that can be used to experimentally unambiguously identify the HEIs.

Our density functional theory calculations were performed using the Quantum Espresso\cite{pwscf} package within the Perdew-Burke-Ernzerhof (PBE) exchange correlation functional\cite{PBE}. Optimized norm-conserving Vanderbilt pseudopotentials\cite{Hamann} were employed with a 70 Ry cut-off. An $18 \times 18 \times 1$ \emph{k}-point grid was used to sample the Brillouin zone. To determine the magnetic ground state, lattice parameters and all atomic positions were fully relaxed until residual forces were less than 1 meV/\AA. To calculate Curie temperature, classical Monte Carlo (MC) simulations were performed on a 200 $\times$ 200 supercell with the Wolff algorithm\cite{Wolff}. Single-shot G$_0$W$_0$ calculations\cite{Hybertsen} were performed for quasiparticle band structure using the Yambo\cite{yambo} code. Energy cut-offs of 60 and 17 Ry, respectively, corresponding to the exchange and correlation parts of the self-energy were employed and a total of 480 bands were used to ensure a gap convergence to within 5 meV. The dielectric functions and exciton energies were obtained by solving the Bethe-Salpeter Equation (BSE)\cite{Rohlfing}.

\begin{figure}[tbp]
\includegraphics[width=0.75\columnwidth]{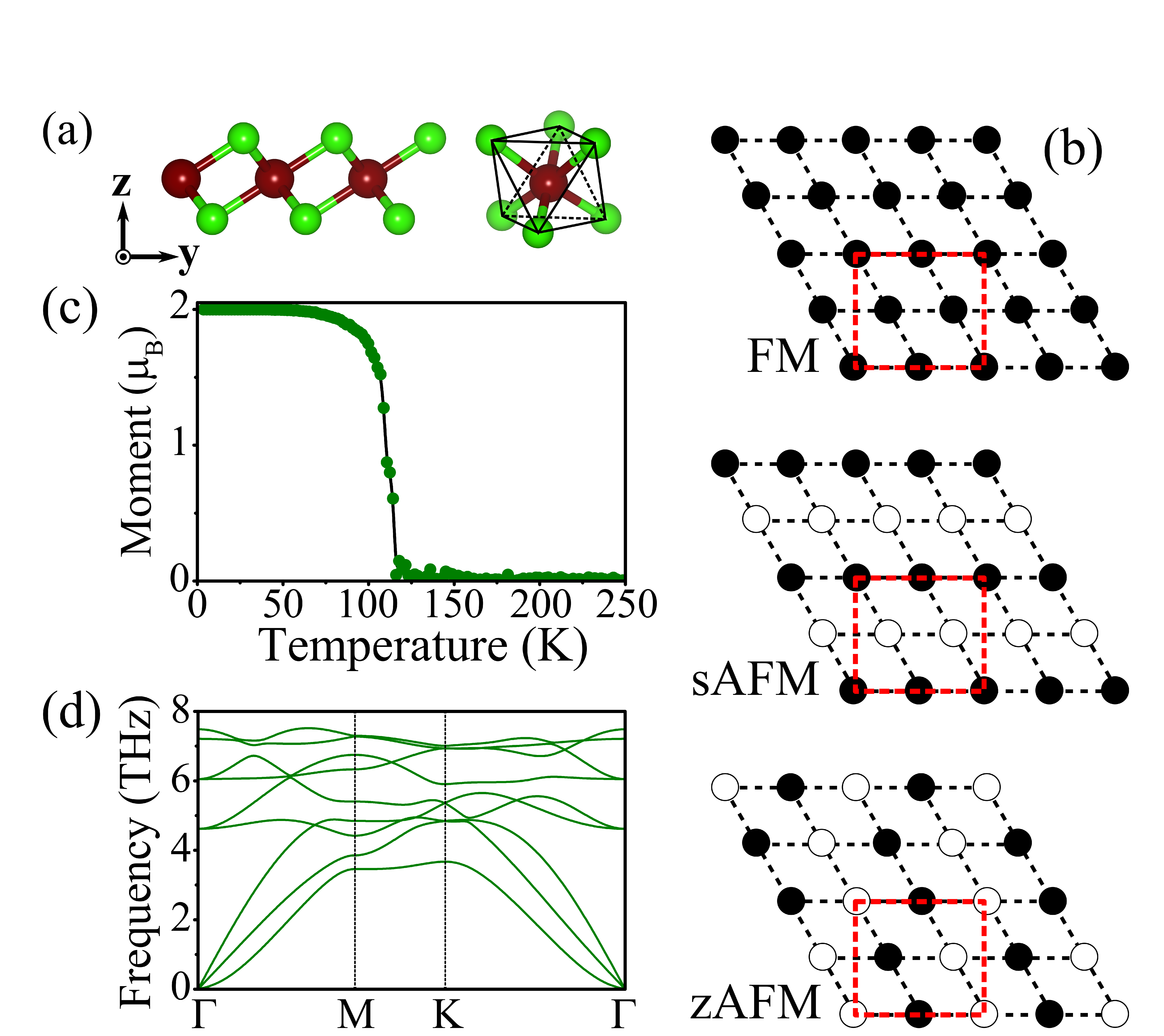}
\caption{\label{fig:fig2} (Color online) (a) Atomic structure (left) and Ni-centered local octahedron (right) of monolayer 1\emph{T}-NiCl$_2$ with Ni (Cl) atoms colored in brown (green). (b) Magnetic phases: FM (top), sAFM (middle), and zAFM (bottom). Open and filled circles are lattice sites on the Ni plane with opposite local moments. Red dashed rectangles are the supercells used for energy calculations. (c) Total magnetic moment versus temperature, showing a Curie temperature of 118 K. (d) Phonon spectrum of the FM (ground) state.}
\end{figure}

The four materials we studied share similar EI properties. Below, we will use NiCl$_2$ as an example to illustrate their common physics. Figure 2(a) shows that its geometry is analogous to 1\emph{T}-MoS$_2$ with a central triangular Ni layer sandwiched by two Cl layers. Four phases were considered, which are non-magnetic (NM), ferromagnetic (FM), stripe- and zigzag-antiferromagnetic (sAFM and zAFM) as illustrated in Fig. 2(b). The FM phase with a total moment 2 $\mu_B$ is energetically favored by 25, 37, and 620 meV per Ni, respectively, to the zAFM, sAFM, and NM phases. Super-exchange is responsible for the long range FM order, as the angles between the Ni-Cl-Ni bond are all around $\sim$$90^\circ$ in accordance with the Goodenough-Kanamori rule\cite{Goodenough,Kanamori}. We have estimated the Curie temperature using a Monte Carlo simulation based on a two-dimensional Ising model with nearest neighbor exchange interactions, which was reported to be more accurate than the mean-field approach.\cite{zhuang} Measuring the moment variations yields a Curie temperature of 118 K [See Fig. 2(c)], which is much higher than that of CrX$_3$\cite{Seyler}. The calculated phonon spectrum in Fig. 2(d) shows no imaginary frequency, which confirms the dynamical stability of the ground-state FM phase. These results on magnetic phase stability agree well with those reported in the literature\cite{Kulish}.

\begin{figure}[tbp]
\includegraphics[width=0.75\columnwidth]{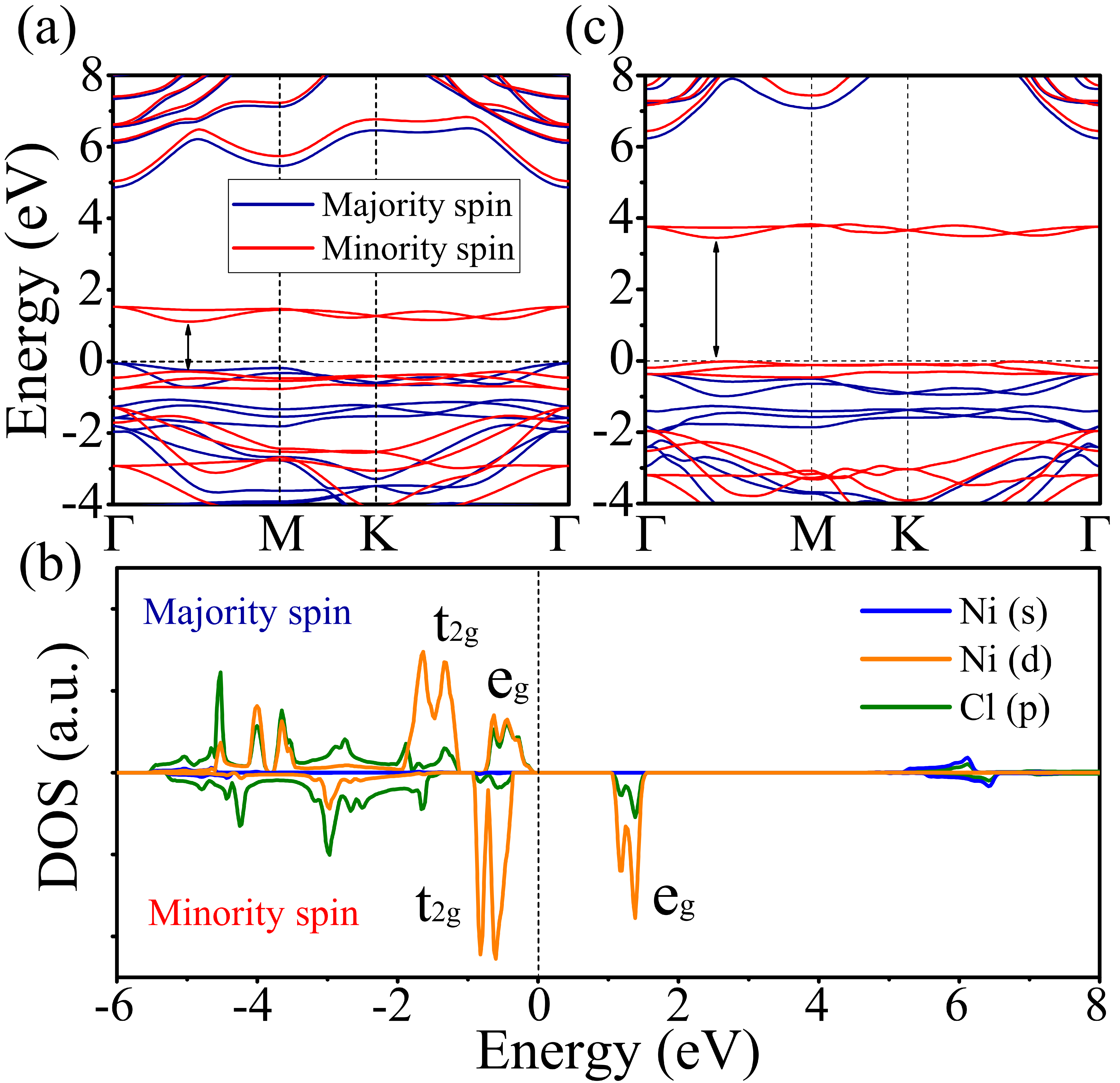}
\caption{\label{fig:fig3} (Color online) Spin-resolved PBE (a) band structure and (b) projected density of states, and (c) the corresponding G$_0$W$_0$ band structure. In (a) and (c), the minority spin gaps are indicated. Energy zero is at the top of valence band.}
\end{figure}

Figure 3(a) depicts the spin-resolved band structure of the FM ground state at the PBE level, showing 1.35- and 4.92-eV charge gaps for the minority and majority spin channels, respectively. While both gaps in Fig. 3(a) are direct, they are located at different \emph{k}-points: namely, the minority spin gap is in the symmetry line between $\Gamma \rm{M}$, whereas the majority spin gap is at $\Gamma$. In the 1\emph{T} structure, each Ni atom is surrounded by six Cl atoms with an approximate octahedral symmetry [See the right of Fig. 2(a)] to result in a $t_{2g}\text{-}e_{g}$ splitting of the $d$ orbitals, as can be seen in the projected density of states in Fig. 3(b). Given that the Ni atom has a $3d^8$ electronic configuration, the lower-lying $t_{2g}$ states are fully occupied while the remaining two electrons fill the majority spin $e_{g}$ states. This yields a total magnetic moment of 2 $\mu_B$.

The orbital symmetry of the bands in Fig. 3(b) suggests a suppressed dielectric screening and hence a reduced dielectric function\cite{Moskvin}, as all the low-energy excitations are parity-forbidden $d$-$d$ and $d$-$s$ transitions\cite{Sugano}. This kind of orbital symmetries decouples $E_b$ from $E_g$, which has been shown to play a critical role in the formation of EI\cite{usEI,usPRL}.

On the other hand, Fig. 3(b) shows that the frontier bands exhibit small dispersions due to the localized nature of Ni $3d$ orbitals. Flat bands are suggestive of strong electron correlations in the system, which may not be properly accounted for by the PBE approach\cite{ourPCCP}. For this reason, we have calculated the more accurate G$_0$W$_0$ quasi-particle band structure, as shown in Fig. 3(c). While on the appearance the bands look similar to those of the PBE, two important points are worthy to note: first, the gaps are increased considerably by almost 2 eV to 3.37 and 6.30 eV, respectively; second, the band structure is qualitatively different from that of the PBE in the sense that the minority, instead of the majority, spin is now responsible for the valence band maximum. To further justify the band characteristics, we performed dynamical mean-filed theory calculations, which is not only based on substantially different approximations from density functional theory but also a powerful method for strongly correlated electrons\cite{DMFT1,DMFT2,DMFT3}. The results shown in Fig. S1 \cite{SI} reveal the same ground-state electronic properties as reported above.

The localized nature of the Ni $3d$ orbitals suggests the formation of tightly-bound Frenkel-like excitons, whose binding energy can be greatly enhanced by an increased electron-hole interaction in two dimension due to the reduced dielectric screening. To this end, we have solved the BSE for the dielectric function and exciton energies. Figure 4(a) shows the imaginary part of the dielectric function, along with the contributions from the two different spins. The two distinct band gaps, namely, those for minority and majority spins (at the energies of 3.37 and 6.30 eV as denoted by red and blue dots), divide the spectrum into three regions. When the excitation energy is less than the $E_g$ of minority spin, the excitations are mainly those associated with $d$-$d$ transitions within the minority spin. When the excitation energy is larger than the $E_g$ of minority spin but smaller than the $E_g$ of majority spin, the excitations are associated with $p$-$d$ transitions within the minority spin. When the excitation energy is larger than the $E_g$ of majority spin, the excitations become much less featured and spin-independent.

Given the symmetry-forbidden transitions at low energies, dark excitons can be expected. Figure 4(b) compares all the excitons (i.e., gray vertical lines) calculated by the BSE with the low-energy absorption spectrum below the $E_g$ of minority spin. Two dark excitons below the $d$-$d$ transitions (denoted by $X_1$ and $X_2$) are noteworthy. In particular, the $X_1$ consists of a ``double-line" structure with an energy splitting of 27 meV. As it turns out, these two excitons share a similar physics, so in the following we will focus our discussion on the $X_1$ state with lower energy.

\begin{figure}[tbp]
\includegraphics[width=0.75\columnwidth]{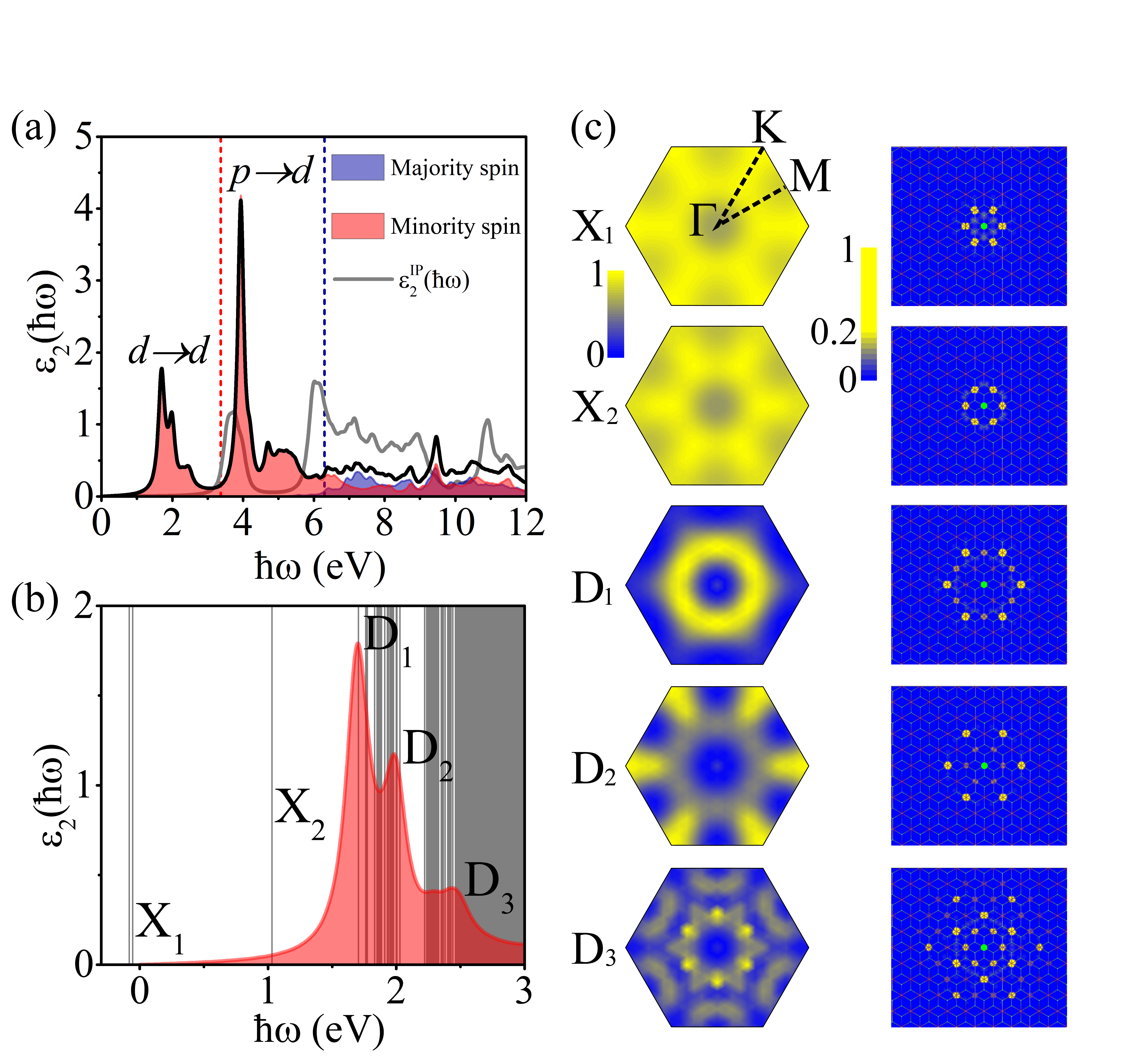}
\caption{\label{fig:fig4} (Color online) (a) Imaginary part of the dielectric function. The black solid line is obtained from the BSE with contributions from individual spins color-coded with red and blue, vertical lines denoting the corresponding $E_g$. The gray solid line is obtained from the independent-particle (IP) approximation, i.e., ignoring electron-hole interactions. (b) Exciton energies (vertical lines), superimposed on the imaginary part of the BSE dielectric function in (a) in the low-energy region. $X_i$ and $D_i$ denote dark excitons and the peak positions of $\varepsilon_2$, which are indicative of bright excitons. (c) Reciprocal-space (left) and real-space (right) exciton wavefunctions modulus. The real-space wavefunctions are highly localized on Ni atoms. To show the fine details, we truncate the wavefunctions modulus at 20\% of their maximum values. In the plots, the hole (the green dot) is fixed at central Ni.}
\end{figure}

For comparison, three optically visible peaks are also denoted by $D_1$, $D_2$, and $D_3$ in Fig. 4(b). These bright excitons are expected to be distinctly different from $X_1$ and $X_2$. To elucidate the differences, Fig. 4(c) (left) plots the reciprocal-space wavefunctions for the $X$ and $D$ series. One may note that these excitons occupy all or most of the Brillouin zone (BZ), which are contrasted to the Wannier-Mott excitons occupying only a small part of the BZ\cite{Elliott}, but are in great resemblance to the Frenkel excitons. Indeed, real-space exciton wavefunctions in Fig. 4(c) (right) reveal that both the $X$ and $D$ excitons are highly localized to within only a few unit cells. Specifically, the structured $X_1$ and $X_2$ reflect the hopping between nearest neighbor Ni atoms but with subtle differences. The structured $D_1$ and $D_2$ reflect, on the other hand, second nearest neighbor hopping, while that of $D_3$ reflects the third nearest neighbor hopping. Generally speaking, the further away the hopping distance between Ni atoms, the further away the electron and hole centers, and hence the smaller the $E_b$. As the exciton radius increases, Fig. 4(c) shows that more Cl states are included in the exciton wavefunction, whose hybridization with the $d$ states of Ni atoms progressively breaks the parity selection rule\cite{Sugano}.

The rich spectral features in Fig. 4(b) are in fact many-body effects due to the nearly flat $d$ bands. To see this, we show in Fig. 4(a) by gray line the independent-particle (IP) dielectric function $\varepsilon_2^{IP}(\hbar\omega)$ \cite{noteforIP}, which is calculated within the single-electron picture. In a startle contrast to the numerous distinguishable peaks in $\varepsilon_2(\hbar\omega)$, $\varepsilon_2^{IP}(\hbar\omega)$ exhibits only a featureless broad peak in the energy range between 3.4 and 4.2 eV, as a reminiscence of the nearly degenerate single-electron $d$ bands in Fig. 3(b).

The Coulomb interaction between 3$d$ electrons is noteworthy. Previous study\cite{Kulish} showed that this can change the relative energies of Ni 3$d$ and Cl 3$p$ orbitals, causing significant $p$-$d$ hybridization and possibly affecting the low-energy excitons. To explore the effect, we conducted hybrid functional (HSE06 \cite{HSE1,HSE2}) and PBE+$U$ \cite{NiO1} calculations. The calculation details and results are given in the Supplemental Material\cite{SI}. Indeed, we find an increased $p$-$d$ hybridization for both two kinds of calculations. However, the essence of low-energy excitons remains largely unchanged and a stronger excitonic instability is manifested by a much more negative $E_t$. Specifically, HSE06@BSE yields an $E_t$ = $-$945 meV, which is one order of magnitude larger than $E_t$ = $-83$ meV by PBE@G$_{0}$W$_{0}$@BSE. Similar trend is revealed by PBE+$U$ method based on physically meaningful $U$ values of 5$\sim$6 eV that can reproduce the experimental spectra of NiO \cite{NiO1,NiO2}. Also, PBE+$U$@BSE yields an $E_t$ = $-$1220 ($-$847) meV corresponding to $U$ = 5 (6) eV, comparable with the HSE06@BSE value of $-945$ meV.

As discussed earlier, a negative $E_t$ implies thermodynamic instability in the minority-spin channel of a band insulator, leading to the formation of HEI, as shown in Fig. 1(c). When this happens, the system is expected to display novel single-spin phenomena associated with a Bose-Einstein condensate\cite{Kohn,Rontani,usEI} in the minority spin. Unlike other EIs being studied to date, however, here the excitonic instability occurs in systems with an $E_g$ up to 3.37 eV, which are insulating under the ambient conditions. This dramatic result is a direct consequence of the strong correlation effect of localized 3$d$ electrons, which causes a significantly-reduced screening and subsequently highly tightly-bound excitons. Many-body flat-band effects have recently attracted considerable attentions, e.g., unconventional superconductivity in magic-angle graphene\cite{Cao}, high-temperature fractional quantum Hall states\cite{Tang} and ferromagnetic ordering\cite{Lin}.

The condensation of spin-singlet excitons in only one of the spin channels is distinctly different from the condensation of excitons in cobalt oxides, which are spin triplet\cite{Kunes2014,Nasu2016}. In particular, the onset of the EI state here is not accompanied by a usual metal-insulator transition. Rather, the emergence of the single-spin-channel EI is closely related to the unique characteristics of the magnetic band structure of the host material. One may attempt to describe such a system using the (extended) Falicov-Kimball model designed for the spin-singlet exciton condensate\cite{Kaneko,Seki2011,Kunes2015}. However, both intra- and inter-band Coulomb interactions may have to be included to properly describe our systems.

\begin{figure}[tbp]
\includegraphics[width=0.75\columnwidth]{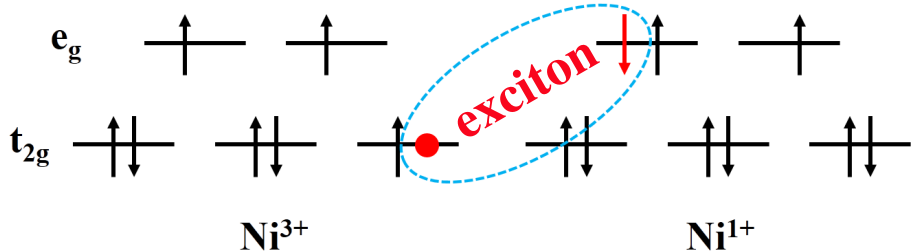}
\caption{\label{fig:fig5} (Color online) A schematic illustration of the formation of $X_1$ exciton. It corresponds to a $|d^8d^8$$>$$\rightarrow|d^7d^9$$>$ excitation. ``$\uparrow$" and ``$\downarrow$" denote occupied $d$-electrons with up and down spin. An $X_1$, as can be seen in the encircled area, is formed by elevating a down-spin electron (red ``$\downarrow$") to the nearest neighbor Ni, leaving a hole (red filled circle), which makes two adjacent Ni atoms differ in their valency (+3 and +1).}
\end{figure}

The formation of excitons, in particular, the $X_1$, alters the magnetic properties from those of a single-particle band picture, as schematically illustrated in Fig. 5. Within the single-electron picture, each Ni atom ($3d^8$) fully occupies the $t_{2g}$ states as well as the majority spin $e_{g}$ states with FM ordering as the ground state. With the exciton $X_1$ state, however, an electron in the minority $t_{2g}$ states could ``permanently" hop to the neighboring Ni $e_g$ states, as the hopping lowers the system energy by forming HEI. This causes the adjacent Ni atoms, on which the exciton resides, differ in their valencies (+3 and +1). Note that none of the excitons is localized on a particular atom. In other words, one can move the holes in the real-space plots in Fig. 4(c) to any Ni atom. This implies that all Ni should have a mixed valency, which is a hallmark of double-exchange mechanism for FM\cite{Zener}, opposed to the super-exchange mechanism discussed earlier within the single-electron band picture.

In a non-magnetic system, it is usually the large electron-hole binding energy ($E_{e\text{-}h}$) that drives the EI formation. In a magnetic system, however, exchange interaction ($E_{ex}$) also contributes. Taking the case in Fig. 5 as an example, the formation of $X_1$ benefits from both a Coulomb attraction between the electron-hole pair and an exchange coupling of the excited electron with three $t_{2g}$ states on the same Ni$^{1+}$ ion.  These energies compete with the excitation energy between $t_{2g}$ and $e_g$ (i.e., $E_{t_{2g}\text{-}e_g}$). Evidently, the exciton stabilization energy $E_{e\text{-}h}+E_{ex}-E_{t_{2g}\text{-}e_g}$ depends on the coupling between local moments. Thus, the strength of the magnetic coupling is responsible for two transition temperatures: one is $T_C^M$ for magnetic phase transition, while the other is $T_C^E$ for the formation of exciton condensate. When $T_C^M < T_C^E$, one can have a situation where the system is non-magnetic in the one-particle spectrum but magnetic in the many-body spectrum.

While all four HEIs in our study share commonality, there are also differences. For example, monolayer 1\emph{T}-CoCl$_2$ and 1\emph{T}-CoBr$_2$ exhibit a spontaneous symmetry-breaking due to one less electron of Co than Ni, which splits the Co $t_{2g}$ states further into a doublet and a singlet. The resulting ``intra-$t_{2g}$" transitions give rise to an excitonic instability in the minority spin. Because of the relatively small energy splitting of $t_{2g}$ for Co, much larger $E_t$s are found for CoCl$_2$ and CoBr$_2$ as compared to those of NiCl$_2$ and NiBr$_2$.

Let us now turn to experimental verification of HEIs. In this regard, optical absorption is a direct tool to differentiate between band and excitonic insulators. In a band insulator, the optical gap is given by the difference between $E_g$ and $E_b^{br}$ where the superscript (br) denotes the first bright exciton, while in an EI, it would be $E_b$, provided that one can ignore many-body excited states\cite{Lu,Du}. Angle-resolved photoemission spectroscopy (ARPES) is another useful tool to observe such a characteristic transition. From a practical point of view, however, the Ni systems with a relatively small $E_t <$ 100 meV might be easier to study than the Co systems for their significantly larger $E_t$. One can also directly measure the magnetic responses: for example in the case when $T_C^M > T_C^E$, while the system remains to be in the FM ground state, the formation of HEI qualitatively alter the magnetic coupling mechanism. Hence, a change in the magneto resistivity and/or magneto optical responses should take place near $T_C^E$.

In summary, using first-principles calculations we identify a new state of matter --- the half excitonic insulators which combines charge gap and magnetism with spontaneous formation of excitons. Excitons are bosons. As such, the results here also blend magnetic degrees of freedom with Bose-Einstein condensation. The many-body physics of the HEIs distinguishes themselves from the physics of band insulators markedly. Hence, we expect the HEIs to offer a great potential in electronic, spintronic, and photonic applications. Moreover, since it is the dark exciton that causes the excitonic instability, low-energy excitations in such an EI may involve transitions only between many-body dark states, opposed to those between one-particle ground state and many-body excited states\cite{CaoT,ZhangXO}. This unique feature not only allows for a direct probing of the dark excitons, which has not been possible before, but also offers the possibility of establishing selection rules (if any) or at least the spectral weights in many-body systems for which the physics is still in its infancy.

\begin{acknowledgments}
Y.L. thanks Yu Liu and Haifeng Song for  fruitful discussions. Work in China was supported by the Basic Science Center Project of NSFC (Grant No. 51788104), the Ministry of Science and Technology of China (Grant No. 2016YFA0301001), the National Natural Science Foundation of China (Grant Nos. 11674071 and 11674188), the Beijing Advanced Innovation Center for Future Chip (ICFC), the Open Research Fund Program of the State Key Laboratory of Low-Dimensional Quantum Physics (NO. KF201702), and the Beijing Institute of Technology Research Fund Program for Young Scholars. Work in the US (SBZ) was supported by US DOE under Grant No. DE-SC0002623. SBZ had been actively engaged in the design and development of the theory, participated in all the discussions and draft of the manuscript.
\end{acknowledgments}


\begin{thebibliography}{90}%
\makeatletter
\bibitem{High} A. A. High, E. E. Novitskaya, L. V. Butov, M. Hanson, and A. C. Gossard, Science \textbf{321}, 229 (2008).

\bibitem{Kohn} D. J\'{e}rome, T. M. Rice, and W. Kohn, Phys. Rev. \textbf{158}, 462 (1967).

\bibitem{Kogar} A. Kogar, S. Vig, M. S. Rak, A. A. Husain, F. Flicker, Y. I. Joe, L. Venema, G. J. MacDougall, T. C. Chiang, E. Fradkin, J. van Wezel, and P. Abbamonte, Science \textbf{358}, 1314 (2017).

\bibitem{Rontani} M. Rontani and L. J. Sham, in \emph{Novel Superfluids}, edited by K. H. Bennemann and J. B. Ketterson (Oxford University Press, Oxford, 2014), vol. 2, p. 423.

\bibitem{Kunes2015} J. Kune\v{s}, J. Phys.: Condens. Matter \textbf{27,} 333201 (2015).

\bibitem{RontaniPRL} M. Rontani and L. J. Sham, Phys. Rev. Lett. \textbf{94,} 186404 (2005).

\bibitem{usEI}  Z. Y. Jiang, Y. C. Li, S. B. Zhang, and W. H. Duan, Phys. Rev. B \textbf{98,} 081408(R) (2018).

\bibitem{HanW} W. Yuan, Q. Zhu, T. Su, Y. Y. Yao, W. Y. Xing, Y. Y. Chen, Y. Ma, X. Lin, J. Shi, R. Shindou, X. C. Xie, W. Han, Sci. Adv. \textbf{4,} eaat1098 (2018).

\bibitem{McGuire} M. A. McGuire, Crystals \textbf{7,} 121 (2017).

\bibitem{pwscf} P. Giannozzi, S. Baroni, N. Bonini, M. Calandra, R. Car, C. Cavazzoni, D. Ceresoli, G. L. Chiarotti, M. Cococcioni, I. Dabo, A. D. Corso, S. de Gironcoli, S. Fabris, G. Fratesi, R. Gebauer, U. Gerstmann, C. Gougoussis, A. Kokalj, M. Lazzeri, L. Martin-Samos, N. Marzari, F. Mauri, R. Mazzarello, S. Paolini, A. Pasquarello, L. Paulatto, C. Sbraccia, S. Scandolo, G. Sclauzero, A. P. Seitsonen, A. Smogunov, P. Umari, and R. M. Wentzcovitch, J. Phys.: Condens. Matter \textbf{21,} 395502 (2009).

\bibitem{PBE} J. P. Perdew, K. Burke, and M. Ernzerhof, Phys. Rev. Lett. \textbf{77,} 3865 (1996).

\bibitem{Hamann} D. R. Hamann, Phys. Rev. B \textbf{88,} 085117 (2013).

\bibitem{Wolff} U. Wolff, Phys. Rev. Lett. \textbf{62,} 361 (1989).

\bibitem{Hybertsen} M. S. Hybertsen and S. G. Louie, Phys. Rev. B \textbf{34,} 5390 (1986).

\bibitem{yambo} A. Marini, C. Hogan, M. Gr\"{u}ning, and D. Varsano, Comput. Phys. Commun. \textbf{180,} 1391 (2009).

\bibitem{Rohlfing} M. Rohlfing and S. G. Louie, Phys. Rev. B \textbf{62,} 4927 (2000).

\bibitem{Goodenough} J. B. Goodenough, Phys. Rev. \textbf{100,} 564 (1955).

\bibitem{Kanamori} J. Kanamori, J. Phys. Chem. Solids \textbf{10,} 87 (1959).

\bibitem{zhuang} H. L. Zhuang, Y. Xie, P. R. C. Kent, and P. Ganesh, Phys. Rev. B \textbf{92,} 035407 (2015).

\bibitem{Seyler} K. L. Seyler, D. Zhong, D. R. Klein, S. Gao, X. Zhang, B. Huang, E. Navarro-Moratalla, L. Yang, D. H. Cobden, M. A. McGuire, W. Yao, D. Xiao, P. Jarillo-Herrero, and X. Xu, Nat. Phys. \textbf{14,} 277 (2018).

\bibitem{Kulish} V. V. Kulish and W. Huang, J. Mater. Chem. C. \textbf{5,} 8734 (2017).

\bibitem{Moskvin} A. S. Moskvin, A. A. Makhnev, L. V. Nomerovannaya, N. N. Loshkareva, and A. M. Balbashov, Phys. Rev. B \textbf{82,} 035106 (2010).

\bibitem{Sugano} S. Sugano, Y. Tanabe, and H. Kamimura, \emph{Multiplets of Transition - Metal Ions in Crystals} (Academic, New York, 1970).

\bibitem{usPRL} Z. Y. Jiang, Z. R. Liu, Y. C. Li, and W. H. Duan, Phys. Rev. Lett. \textbf{118,} 266401 (2017).

\bibitem{ourPCCP} H. X. Tan, Y. C. Li, S. B, Zhang, and W. H. Duan, Phys. Chem. Chem. Phys. \textbf{20,} 18844 (2018).

\bibitem{DMFT1} G. Kotliar, S. Y. Savrasov, K. Haule, V. S. Oudovenko, O. Parcollet, and C. A. Marianetti, Rev. Mod. Phys. \textbf{78,} 865 (2006).

\bibitem{DMFT2}	A. Georges, G. Kotliar, W. Krauth and M. J. Rozenberg, Rev. Mod. Phys. \textbf{68,} 13 (1996).

\bibitem{DMFT3} http://hauleweb.rutgers.edu/database$\textunderscore$w2k/.

\bibitem{SI} See Supplemental Material for more details of the dynamical mean-filed theory calculations, the hybrid functional calculations and PBE+$U$ calculations.

\bibitem{Elliott} R. J. Elliott, Phys. Rev. B \textbf{108,} 1384 (1957).

\bibitem{noteforIP} Neglecting the local field effect within the random phase approximation gives the independent-particle
dielectric function. See Ref. \onlinecite{usEI} for more details.

\bibitem{HSE1} J. Heyd, G. E. Scuseria, and M. Ernzerhof, J. Chem. Phys. \textbf{118,} 8207 (2003).

\bibitem{HSE2} J. Heyd, G. E. Scuseria, and M. Ernzerhof, J. Chem. Phys. \textbf{124,} 219906 (2006).

\bibitem{NiO1}	S. L. Dudarev, G. A. Botton, S. Y. Savrasov, C. J. Humphreys, and A. P. Sutton, Phys. Rev. B \textbf{57,} 1505 (1998).

\bibitem{NiO2}	O. Bengone, M. Alouani, P. Bl\"{o}chl, and J. Hugel, Phys. Rev. B \textbf{62,} 16392 (2000).

\bibitem{Cao} Y. Cao, V. Fatemi, S. Fang, K. Watanabe, T. Taniguchi, E. Kaxiras, and P. Jarillo-Herrero, Nature \textbf{556}, 7699 (2018).

\bibitem{Tang} E. Tang, J. -W Mei, and X. -G. Wen, Phys. Rev. Lett. \textbf{106,} 236802 (2011).

\bibitem{Lin} Z. Y. Lin, J.-H. Choi, Q. Zhang, W. Qin, S. Yi, P. D. Wang, L. Li, Y. F. Wang, H. Zhang, Z. Sun, L. M. Wei, S. B. Zhang, T. F. Guo, Q. Y Lu, J.-H. Cho, C. G. Zeng, and Z. Y. Zhang, Phys. Rev. Lett. \textbf{121,} 096401 (2018).

\bibitem{Kunes2014} J. Kune\v{s}, and P. Augustinsk\'{y}, Phys. Rev. B \textbf{90,} 235112 (2014).

\bibitem{Nasu2016} J. Nasu, T. Watanabe, M. Naka, and S. Ishihara, Phys. Rev. B \textbf{93,} 205136 (2016).

\bibitem{Seki2011} K. Seki, R. Eder, and Y. Ohta, Phys. Rev. B \textbf{84,} 245106 (2011).

\bibitem{Kaneko} T. Kaneko and Y. Ohta, Phys. Rev. B \textbf{94,} 125127 (2016).

\bibitem{Zener} C. Zener, Phys. Rev. \textbf{82,} 403 (1951).

\bibitem{Lu} Y. F. Lu, H. Kono, T. I. Larkin, A. W. Rost, T. Takayama, A. V. Boris, B. Keimer, and H. Takagi,  Nat. Commun. \textbf{8,} 14408 (2017).

\bibitem{Du} L. J. Du, X. W. Li, W. K. Lou, G. Sullivan, K. Chang, J. Kono, and R. R. Du, Nat. Commun. \textbf{8,} 1971 (2017).

\bibitem{CaoT} T. Cao, M. Wu, and S. G. Louie, Phys. Rev. Lett. \textbf{120,} 087402 (2018).

\bibitem{ZhangXO} X. O. Zhang, W.-Y. Shan, and D. Xiao, Phys. Rev. Lett. \textbf{120,} 077401 (2018).
\end{thebibliography}
\end{document}